\documentclass[reprint, prl, aps, showpacs]{revtex4-1}

\usepackage{graphicx}
\usepackage{amsmath,amssymb}
\usepackage[usenames,dvipsnames]{color}
\usepackage[colorlinks=true, citecolor=blue, linkcolor=WildStrawberry]{hyperref}

\newcommand{\drm}{{\rm d}}
\newcommand{\irm}{{\rm i}}
\newcommand{\beq}{\begin{equation}}
\newcommand{\eeq}{\end{equation}}
\newcommand{\bdm}{\begin{displaymath}}
\newcommand{\edm}{\end{displaymath}}

\DeclareFontFamily{OT1}{pzc}{}
\DeclareFontShape{OT1}{pzc}{m}{it}{<-> s * [1.10] pzcmi7t}{}
\DeclareMathAlphabet{\mathpzc}{OT1}{pzc}{m}{it}

\graphicspath{{./plots/}}
\begin{document}

\title{Upper Limit on a Stochastic Background of Gravitational Waves from Seismic Measurements in the Range 0.05\,Hz to 1\,Hz}

\author{Michael Coughlin}
\affiliation{Department of Physics, Harvard University, Cambridge, MA 02138, USA}

\author{Jan Harms}
\affiliation{INFN, Sezione di Firenze, Sesto Fiorentino, 50019, Italy}

\begin{abstract}
In this paper, we present an upper limit of  $\Omega_{\rm GW}<1.2\times 10^{8}$ on an isotropic stochastic gravitational-wave (GW) background integrated over a year in the frequency range 0.05\,Hz -- 1\,Hz, which improves current upper limits from high-precision laboratory experiments by about 9 orders of magnitude. The limit is obtained using the response of Earth itself to GWs via a free-surface effect described more than 40 years ago by F.~J.~Dyson. The response was measured by a global network of broadband seismometers selected to maximize the sensitivity.
\end{abstract}

\maketitle
The idea to use Earth itself as response body to GWs is occasionally discussed among GW scientists, but quickly refuted since back-of-the-envelop calculations show that achievable strain sensitivities lie far below what is commonly considered as scientifically interesting. In fact, past attempts to find GW signals of known frequency from pulsars in seismic data were either unsuccessful \cite{WiPr1969,MNS1972} or even led to false detection claims \cite{SaMe1972}. As illuminating as these publications are from today's perspective, it will be shown in this paper that analyzing data from a network of modern global broadband seismometers with near optimal search pipelines can nonetheless lead to upper limits on GW amplitudes that beat previous high-precision gravity-strain measurements in certain frequency bands by many orders of magnitude. More specifically, we set a new upper limit on a frequency-independent energy density of a stationary stochastic GW background in the frequency range 0.05\,Hz -- 1\,Hz that is about 9 orders of magnitude below the previous upper limit.  

The response mechanism exploited here was first described by \citet{Dys1969}. In the following, we repeat the most important part of his calculation mainly to present it in a modern form. Dyson derives a boundary condition at a free flat surface that links surface displacement associated with seismic waves to the strain tensor of a GW. This equation can readily be written as
\beq
\begin{split}
&\lambda(\nabla\cdot\vec\xi(\vec r\,,t))\vec e_z+\mu ((\vec e_z\cdot\nabla)\vec\xi(\vec r\,,t)+\nabla(\vec e_z\cdot\vec\xi(\vec r\,,t)))\\
&\qquad\qquad=\mu\,\vec e_z^\top\cdot h(\vec r\,,t)
\label{eq:Dyson}
\end{split}
\eeq
Here $\lambda,\,\mu$ are the Lam\'e constants, $\vec\xi(\vec r\,,t)$ ground displacement, $\vec e_z$ the normal vector to the surface, and $h(\vec r\,,t)$ the spatial part of the GW strain tensor (i.~e.~a $3\times 3$ matrix). This equation can be used to derive a GW response measured in horizontal and vertical surface displacement, but since seismic noise in vertical direction is typically weaker than in horizontal direction, we will focus on vertical displacement $\xi_z$ from here on.

In this case, the response to GWs can be obtained by first projecting equation (\ref{eq:Dyson}) onto the $z$-direction:
\beq
\lambda(\nabla\cdot\vec\xi(\vec r\,,t))+2 \mu\partial_z\xi_z(\vec r\,,t)=\mu\,\vec e_z^\top\cdot h(\vec r\,,t)\cdot \vec e_z
\label{eq:vertical}
\eeq
As pointed out in \citep{Dys1969}, boundary conditions demand that the horizontal wavevector of the GW and the generated seismic waves are the same. This means that since GWs travel at much higher speed than seismic waves, seismic waves generated by a GW propagate almost vertically with respect to the surface. In this case, all vertical seismic displacement is produced by longitudinal (compressional) waves, and all horizontal displacement by transverse (shear) waves. With these conclusions, we can solve equation (\ref{eq:vertical}) via Fourier transform with respect to $\vec r$ and $t$, using the following approximations 
\beq
\begin{split}
\nabla\cdot\vec\xi(\vec r\,,t)&\rightarrow \irm\vec k\cdot\vec\xi(\vec k\,,\omega)\approx\irm\,\omega/\alpha\,\xi_z(\vec k\,,\omega)\\
\partial_z\xi_z(\vec r\,,t)&\rightarrow \irm k_z\xi_z(\vec k\,,\omega)\approx\irm\,\omega/\alpha\,\xi_z(\vec k\,,\omega)
\end{split}
\eeq
where $\alpha$ is the speed of compressional waves, and $\vec k,\omega$ are the Fourier frequencies associated with $\vec r,\,t$. Transforming back into $\vec r,\,t$ space, equation (\ref{eq:vertical}) finally simplifies to
\beq
\dot\xi_z(\vec r\,,t)\approx-\frac{\beta^2}{\alpha}\vec e_z^\top\cdot h(\vec r\,,t)\cdot \vec e_z
\label{eq:response}
\eeq
where $\beta$ is the speed of shear waves. The fact that the natural readout variable is ground velocity rather than ground displacement simplifies the analysis as it is directly proportional to the data output by most commercial broadband seismometers, and also seismic noise has favorable properties in these units as shown below.

Applying equation (\ref{eq:response}) in real-world GW searches requires further assumptions and simplifications. First, seismic waves generated by a GW at one location can travel to the other side of the Earth interfering with counter-propagating seismic waves from the same GW and thereby modifying the GW response measured as surface displacement. However, above 0.05\,Hz, seismic waves that have passed Earth have significantly smaller amplitude, which means that the flat free surface GW response is a good approximation to a more refined model that also takes Earth's spherical shape into account \cite{WeBl1965}. More significantly, systematic errors need to be considered with respect to the calibration of data into GW strain according to equation (\ref{eq:response}). The first calibration step is to convert raw data of seismometers into ground velocity. This is relatively easy to achieve in the targeted frequency range and it has also been confirmed in many dedicated experiments that relative calibration errors of broadband seismometers between 0.05\,Hz -- 1\,Hz lie well below 0.1 \cite{VeEA2009}. In addition, spectral histograms have been calculated for a full year to exclude that any of the seismic stations are subject to major technical issues.

The second calibration step from ground velocity to GW strain leads to the dominant systematic error. For an accurate calibration, one needs global surface maps of compressional and shear wave speeds $\alpha,\,\beta$, which are not directly available. They can however be estimated from other parameters.
The most accurate method that we found is to use the Poisson's ratio $\nu$ in combination with a global map of Rayleigh-wave phase velocities $c_{\rm R}$. A global map of the Poisson's ratio was not available, but its value is constrained by measurements to be in the range $\nu\in[0.25,0.29]$ \cite{ZaAm1995}. Therefore, a mean value of $\nu_0=0.27$ is used and an estimate of the calibration error related to global variations $\Delta\nu$ of the Poisson's ratio is obtained from
\beq
\beta^2/\alpha\approx 0.5682c_{\rm R} \cdot(1-1.5377\Delta\nu)
\eeq
The phase velocity map of $c_{\rm R}$ used in this paper was published by \citet{Ek2011}. Since Rayleigh waves show significant dispersion, the maps were evaluated at several frequencies up to $0.04\,$Hz. Below $0.04\,$Hz, dispersion leads to minor relative changes in velocity of less than 0.1 \cite{TrWo1995,Ek2011}. However, Rayleigh-wave velocities decrease rapidly towards higher frequencies \cite{ChMo1975,ChYe1997}. We exploit this general trend to minimize the calibration error by linearly interpolating between $c_{\rm R}$ at 0.05\,Hz to $0.5c_{\rm R}$ at 0.5\,Hz. We estimate the remaining relative calibration error associated with wave dispersion to be about 0.3. Adding this error in quadrature to the relative calibration error associated with variations in Poisson's ratio, which is 0.03, and to the calibration error of the seismometer response to ground motion, leads to an overall calibration error of 0.32. 

Our search for a stochastic GW background is based on the correlation of data from pairs of detectors. An upper limit on a GW energy density $\Omega_{\rm GW}$ is obtained from point estimates of
\beq
Y=2\int\limits_0^\infty\drm f\, \Re[\tilde s_1^*(f)\tilde s_2(f)]\tilde Q_{12}(f)
\label{eq:point}
\eeq
with $\langle Y\rangle=\Omega_{\rm GW}$ provided that noise contributing to the data $\tilde s_1(f),\,\tilde s_2(f)$ is perfectly uncorrelated between seismometers. The signal-to-noise ratio (SNR) can be enhanced by filtering the data \cite{Chr1992,AlRo1999}. The optimal filter spectrum $\tilde Q_{12}(f)$ depends on the so-called overlap-reduction function (ORF) $\gamma_{12}(f)$ \cite{Fla1993}, the noise spectral densities $S_1(f),\,S_2(f)$ of the two seismometers, and also takes into account the relation between the GW spectral density and $\Omega_{\rm GW}$:
\beq
\tilde Q_{12}(f)=\mathcal{N}\frac{\gamma_{12}(f)}{f^3S_1(f)S_2(f)}
\eeq
where $\mathcal{N}$ is a normalization constant \cite{AbEA2012s}. In this form, the filter is optimized for a frequency independent energy density $\Omega_{\rm GW}$. The ORF incorporates the dependence of the optimal filter on the relative positions and alignments of detector pairs. In the case of the free-surface GW response measured in vertical surface velocity, the alignment is fully determined by the detector location. 
\begin{figure}[t]
\centerline{\includegraphics[width=0.95\columnwidth]{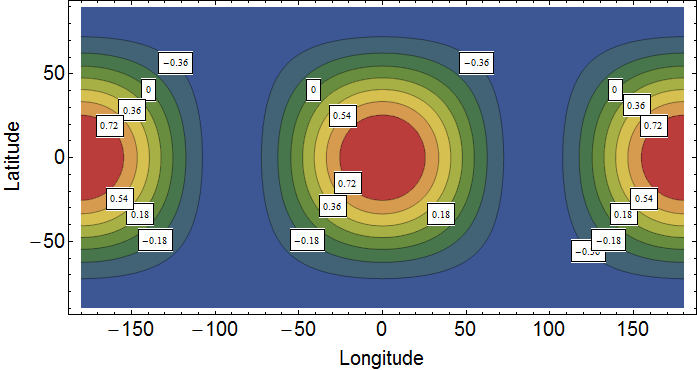}}
\caption{ORF at frequency $f=0.5\,$Hz between seismometers as a function of the latitude and longitude of seismometer 2. Seismometer 1 is located at $\lambda_1=\phi_1=0$.}
\label{fig:ORF}
\end{figure}
Therefore, the ORF is simply a function of the seismometer positions, which is shown in Fig.~\ref{fig:ORF} as a function of the latitude $\lambda_2$ and longitude $\phi_2$ of seismometer 2 with the seismometer 1 at $\lambda_1=\phi_1=0$. For the free-surface response of eq.~(\ref{eq:response}), an explicit expression of the ORF can be obtained:
\beq
\begin{split}
\gamma_{12}(f)&=\frac{15((3-\Phi^2)\sin(\Phi)-3\Phi\cos(\Phi))(1+3\cos(2\delta))}{4\Phi^5}\\
\Phi&\equiv \frac{4\pi f R_\oplus}{c}\sin(\delta/2)\\ \sin^2(\delta/2)&=\sin^2(\Delta\lambda/2)+\cos(\lambda_1)\cos(\lambda_2)\sin^2(\Delta\phi/2)
\end{split}
\eeq
where $f$ is the frequency, $c$ the speed of light, $R_\oplus$ Earth's radius, and $\Delta\lambda=\lambda_2-\lambda_1,\,\Delta\phi=\phi_2-\phi_1$. Here the result is given in terms of the angle $\delta$ subtended by the great circle between the two seismometers, and $\Phi$ is the phase accumulated by a GW that propagates along the line connecting the two seismometers.

Ideally, the locations of seismometers forming a correlation pair should be chosen to maximize the ORF. This is obviously the case for seismometers close to each other (coincident seismometers have $\gamma_{12}(f)=1$). However, choosing seismometers close to each other means that seismic noise is highly correlated, which strongly limits the SNR of the search for a stochastic GW signal. As can be seen in Fig.~\ref{fig:ORF}, the ORF of a pair of antipodal stations is also high, $\rm \gamma_{12}(0.5\,Hz)\approx 0.996$, so that GW signals in the two seismometers are highly correlated, and at the same time, one can expect a great reduction in correlation of seismic noise. Therefore, as a first step, we selected seismometers that form antipodal pairs.

It is found that among seismometers forming antipodal pairs, many still show high seismic correlations. These correlations are generated by teleseismic events that produce significant ground motion on the entire globe. For this reason, times of strong ground motion are excluded in our analysis: two hours following earthquakes with magnitude $M>6$ and a full day after earthquakes with $M>8$. This greatly reduces seismic correlation between stations, but residual correlations can still be significant (i.~e.~much greater than the statistical error). Therefore, instead of using all available seismometer pairs, we select pairs with the lowest seismic correlation observed over an entire year.
\begin{figure}[ht!]
\centerline{\hspace*{-0.4cm}\includegraphics[width=1.2\columnwidth]{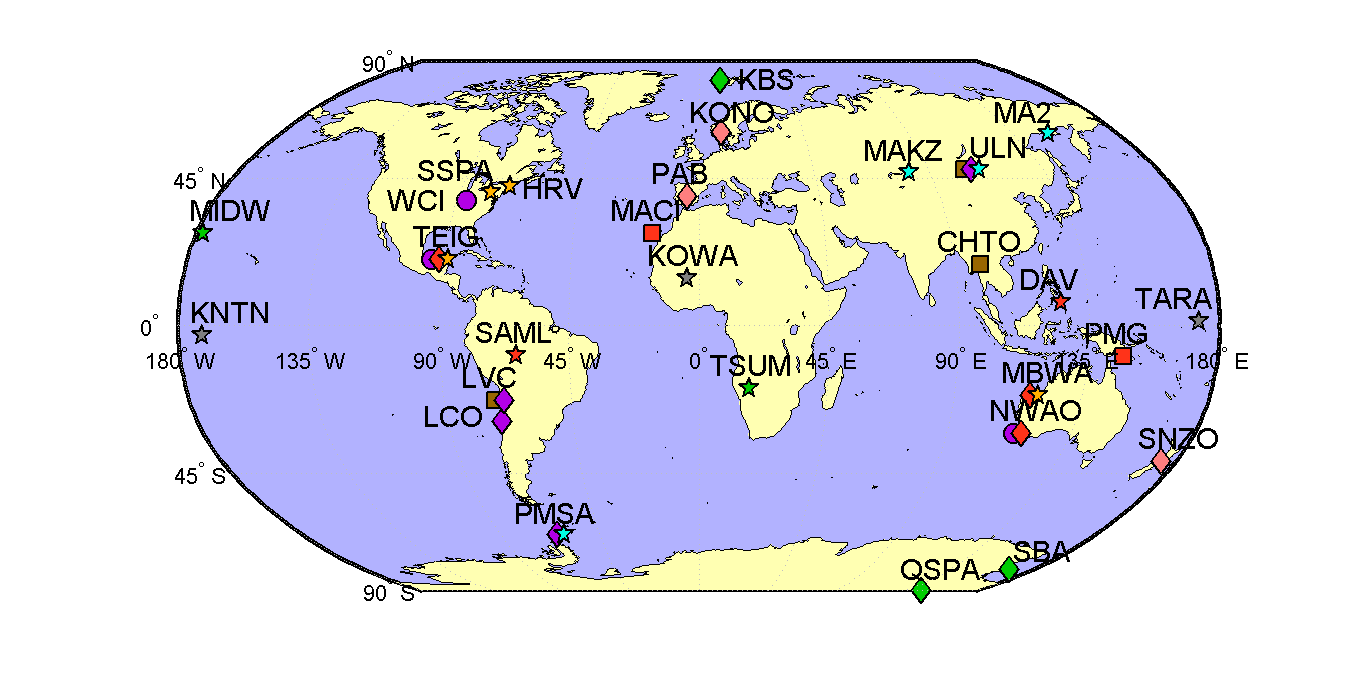}}
\caption{Location of seismometers used in this study. Markers are identical for seismometers that form an antipodal pair. Overlapping markers belong to the same seismic station.}
\label{fig:map}
\end{figure}
The final selection of 20 seismometer pairs is shown in Fig.~\ref{fig:map}. Seismic data from 2012 were used for this study provided by the Data Management Center of the Incorporated Research Institutions for Seismology (IRIS). These stations were equipped with STS-1, STS-2, KS-36000-I, or KS-54000 broadband instruments. Their amplitude and phase response are approximately flat in the frequency range 0.05\,Hz -- 1\,Hz. 

\begin{figure}[ht!]
\centerline{\includegraphics[width=0.9\columnwidth]{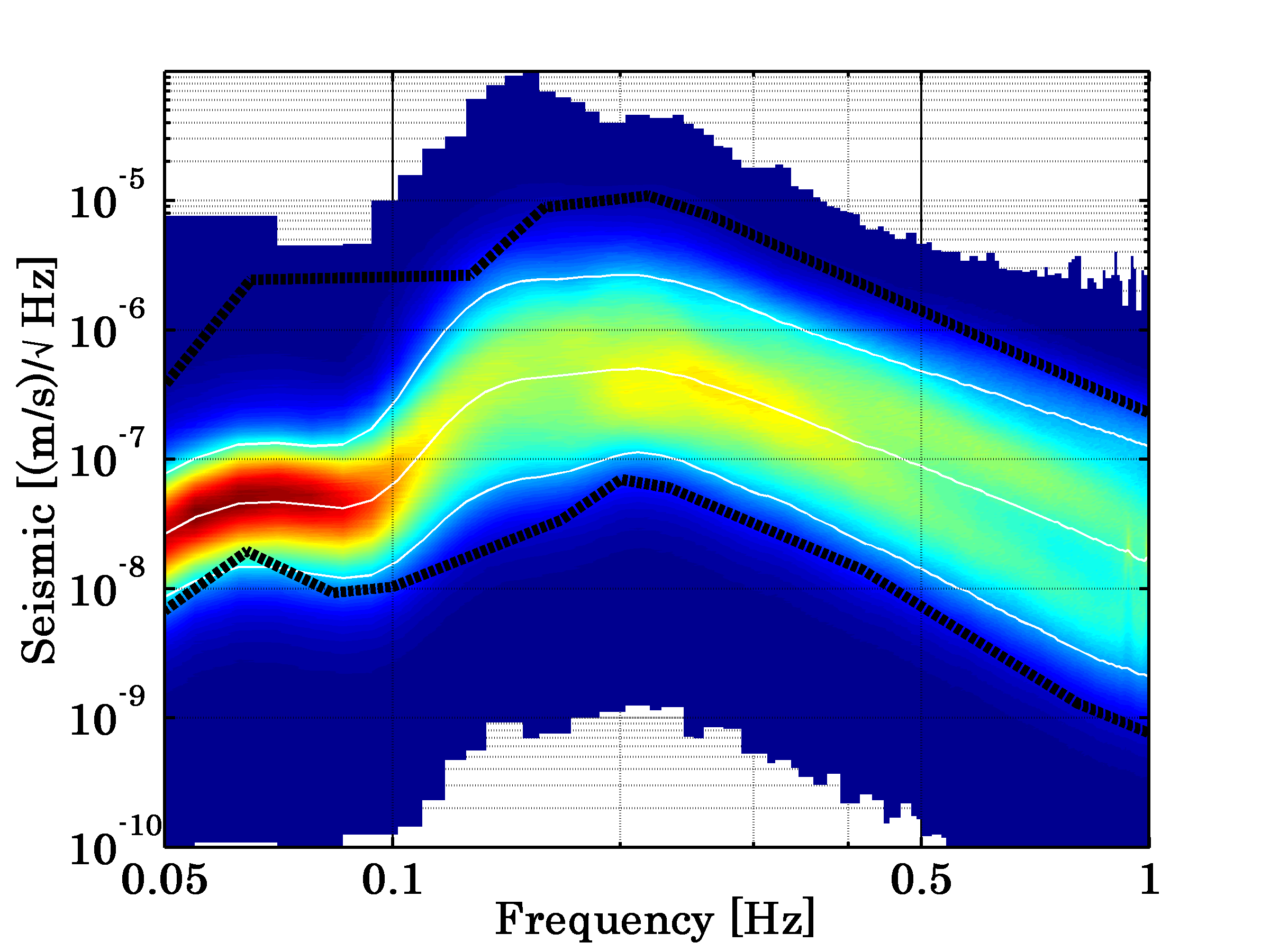}}
\caption{Combined histogram of 128\,s seismic spectra from all stations used in our analysis accumulated over one year excluding vetoed data stretches. The three white curves are the 10th, 50th and 90th percentile of the distribution, whereas the two black lines represent the global low-noise and high-noise models as defined in \cite{Pet1993}.}
\label{fig:histoSEIS}
\end{figure}

As explained in the following, vetoes were applied to data stretches according to different criteria with the goal to restore stationarity of seismic data. The vetoed data are excluded from any of the presented results. As a first step, we give a simple characterization of the seismic data in terms of observed seismic spectra. Measuring a seismic spectrum every 128\,s for each seismometer used in the analysis, and combining all these spectra into one histogram, one obtains the result shown in Fig.~\ref{fig:histoSEIS}.
The histogram does not only provide a complete view on the instrument sensitivity (limited by ambient seismic noise), but was also used to identify stations subject to strong disturbances (either by operators that made changes to the system, or by unknown local seismic events). These stations were further analyzed in time-frequency plots and the affected data stretches were vetoed for further analyses. The resulting distribution of seismic spectra extends well beyond the global low- and high-noise models that are shown as black curves, but the 10th and 90th percentiles lie mostly within this range, which are shown as white curves together with the median consistent with modern definitions of global noise models \cite{CoHa2012b}. 

Another veto was applied based on time series to improve stationarity of the data. What is meant by stationarity of seismic data needs clarification. Any ground motion in the frequency range 0.05\,Hz -- 1\,Hz is produced by some kind of event, be it a storm, earthquake, anthropogenic noise, etc. The distribution of seismic displacement observed over long periods of time and therefore summing the contribution of many thousands of these events is what can be tested for stationarity. It follows that the observation time is an important parameter in the assessment of stationarity. For example, there can be stations in seismically active regions where a background of $M<4$ earthquakes follows a stationary distribution over the course of a year. Since we are ultimately interested in the stationarity of the distribution of $\Omega_{\rm GW}$ point estimates and associated errors, we used the distribution of one of these, the errors, to define a non-stationarity veto of data. This was done by calculating the histogram of errors for each month and station based on 100\,s Hann windowed data segments with 50\% overlap. Data that led to significant variations of these histograms from month to month were vetoed. Ultimately, these vetoes were exclusively related to strong events that contributed to the high-energy tail of the distribution.

After applying all vetoes, which amounted to excluding up to 5\% of data, the resulting combined upper limit using all seismometer pairs, and the upper limit obtained from the single best pair including calibration errors are 
\beq
\Omega^{\rm tot}_{\rm GW}<1.2\times 10^{8},\,\Omega^{\rm sgl}_{\rm GW}<2.7\times 10^{8}
\eeq
We assumed a value $H_0=67.8\,\rm km/s/Mpc$ of the Hubble constant \cite{AdEA2013}. Using $S_{\rm GW}(f) = 3H_0^2\Omega_{\rm GW}/(10\pi^2 f^3)$, this translates into a strain sensitivity of about $1.3\times 10^{-13}\,\rm Hz^{-1/2}$ at 0.1\,Hz. The strongest conceivable upper limit on a stochastic background in the frequency range 0.05\,Hz -- 1\,Hz set by a single seismometer pair can be calculated by assuming that seismic noise is uncorrelated between seismometers and stationary with spectrum given by the global low-noise model, and seismic data is perfectly calibrated into units of GW strain:
\beq
\Omega^{\rm opt}_{\rm GW}<2.4\times 10^7\left(\frac{\rm 1\,yr}{T}\right)^{1/2}\left(\frac{1.9\,{\rm km/s}}{\beta^2/\alpha}\right)^2
\label{eq:optimum}
\eeq
where $T$ is the correlation time, and here seismic speeds are assumed to be frequency independent.

The previous upper limit $\Omega_{\rm GW}< 4.3\times 10^{17}$ was set by a high-precision gravity strain measurement \cite{IsEA2011}. The main reasons for which it was possible to improve this upper limit by so many orders of magnitude are that only modest seismic isolation has been achieved at low frequencies so far. At the same time, the effective baseline of the free-surface response corresponds to the length of seismic shear waves in the range 0.05\,Hz to 1\,Hz, which is about 4--5 orders of magnitude larger than the baseline realized in past experiments. Finally, the global network of broadband seismometers records data reliably, with some stations providing data over periods of several years allowing us to carry out long integrations of the strain signal. 

As can be concluded from eq.~(\ref{eq:optimum}), future improvements on the upper limit by more than an order of magnitude using seismic measurements should not be expected, but it may be possible to achieve a better result by (1) identifying seismometer pairs with unusually low ambient seismic noise at both sites (Moon?), (2) integrating for much longer than 1\,yr, (3) finding locations where the fraction $\beta^2/\alpha$ has an unusually high value, (4) optimizing data selection so that a larger number of seismometer pairs significantly contribute to lowering the upper limit, or (5) narrowing the band of the search and make use of structural resonances in the strain response \cite{Jen1979}.

\section{Acknowledgments}
We are grateful for the constructive comments received by Prof.~Vuk Mandic on an early version of the paper. MC was supported by the National Science Foundation Graduate Research Fellowship
Program, under NSF grant number DGE 1144152. The stochastic GW search has been carried out using the MatApps software available at \url{https://www.lsc-group.phys.uwm.edu/daswg/projects/matapps.html}. Seismic data were downloaded from servers of the IRIS DMC (\url{http://www.iris.edu/SeismiQuery/timeseries.htm}).

\raggedright
\bibliography{references}

\end{document}